# Enhancing Radiographic Disease Detection with MetaCheX, a Context-Aware Multimodal Model

*All authors contributed equally to this work.

Nathan He*
Math and Science Academy
Ocean Lakes High School
Virginia Beach, USA
nathanh43511@gmail.com

Cody Chen*
Los Gatos High School
Los Gatos, USA
codysc@hotmail.com

*Abstract*— Existing deep learning models for chest radiology often neglect patient metadata, limiting diagnostic accuracy and fairness. To bridge this gap, we introduce MetaCheX, a novel multimodal framework that integrates chest X-ray images with structured patient metadata to replicate clinical decision-making. Our approach combines a convolutional neural network (CNN) backbone with metadata processed by a multilayer perceptron through a shared classifier. Evaluated on the CheXpert Plus dataset, MetaCheX consistently outperformed radiograph-only baseline models across multiple CNN architectures. By integrating metadata, the overall diagnostic accuracy was significantly improved, measured by an increase in AUROC. The results of this study demonstrate that metadata reduces algorithmic bias and enhances model generalizability across diverse patient populations. MetaCheX advances clinical artificial intelligence toward robust, context-aware radiographic disease detection.

## I. Introduction

The integration of artificial intelligence (AI) into clinical practices has emerged as a transformative tool across multiple fields of healthcare, especially in diagnostic imaging. AI-based systems have demonstrated significant promise in enhancing the interpretation of medical images, such as X-rays, CT scans, and MRIs, by improving diagnostic accuracy, reducing human error, and streamlining clinical operations. Among medical imaging techniques, chest radiography holds significant clinical value, with over 3 billion chest X-rays (CXRs) performed each year [1]. Because of its rapid acquisition, low radiation dose, and affordability, chest radiography plays a critical role in the timely evaluation, diagnosis, and monitoring of thoracic diseases across diverse healthcare settings.

Chest X-rays are crucial in diagnosing a wide array of diseases such as pneumonia, pulmonary edema, pleural effusion, pneumothorax, and cardiomegaly. They are commonly employed across inpatient, outpatient, and emergency settings, helping clinicians assess patient conditions and determine treatment priorities. Especially useful in low-resource environments where advanced imaging techniques like computed tomography (CT) are not readily accessible, chest X-rays often remain the primary diagnostic imaging method. However, the interpretation of CXRs can be challenging due to their two-dimensional nature, which can obscure abnormalities. Furthermore, because of the variability in radiologist readings, there may be inconsistent or delayed diagnoses.

To address these limitations, deep learning methods have been increasingly applied to automate chest X-ray interpretation. Convolutional neural networks (CNNs) have achieved strong performance by learning hierarchical representations of patterns directly from pixel data. Notably, CheXNet [2] demonstrated strong performance as a binary classification model for pneumonia detection, surpassing practicing radiologists using deep convolutional neural networks, while CheXpert expanded this approach to multi-label classification of fourteen thoracic diseases while considering uncertainty in its labels [3]. Other frameworks have leveraged advanced architectures, even achieving high performance through an ensemble of various CNNs. Deep CNNs, such as those created by Mabrouk et al. had significantly higher AUROC and F1-scores for pneumonia detection compared to respective individual CNNs [4]. Newer models also combine visual and textual modalities, which show significant benefits, particularly in handling unlabeled or mislabeled datasets. A noteworthy vision-language model in chest radiography is ConVIRT, which inputs image and report pairs [5]. Learned features from both modalities were used for classification tasks and zero-shot retrieval tasks, which improved performance, particularly on datasets that lack clear labels [5]. However, a common limitation of these studies is the predominant focus on the radiograph image alone, rather than considering patient-specific metadata such as age, sex, or clinical history.

Patient metadata, such as age, sex, race, and body mass index (BMI), provides valuable context that can improve diagnostic accuracy. For example, the largest demographics of those with pneumonia are adults who are 65 years and older and children who are 5 years and younger due to a compromised immune system [6], the male sex [7], non-Hispanic black individuals [8], and individuals with a high BMI [9]. Incorporating such metadata can also reduce bias, improve model calibration, and ensure fairness across diverse patient groups. The deliberate omission of metadata in many existing models restricts the machine learning model's ability to incorporate contextual information, which would otherwise be routinely considered by radiologists, limiting diagnostic accuracy and model generalizability, especially for underrepresented populations. Furthermore, the majority of existing datasets for chest radiology contain significant imbalances with disease type, as well as demographic information. Our work addresses this gap by explicitly integrating metadata with imaging features to enhance disease diagnosis and account for patient-specific conditions.

## II. MATERIALS AND METHODS

### A. Dataset and Labeling

Our model utilizes the CheXpert Plus dataset, an extension of the original CheXpert dataset released by Stanford University, which contains 223,228 unique pairs of radiology reports and chest X-rays [10]. Because of its large size, the performance of our model was enhanced in comparison to using smaller datasets released before CheXpert Plus. A total of 14 pathology labels were generated from the radiology reports, including atelectasis, cardiomegaly, consolidation, edema, enlarged cardio mediastinum, fracture, lung lesion, lung opacity, pleural effusion, pleural other, pneumonia, pneumothorax, support devices, and no finding [10]. For each of these diseases, the label was either positive, uncertain, negative, or not mentioned [10]. We treated uncertain labels as negative and ingored the not mentioned labels during the loss computation as was done in the original CheXpert framework in order to reduce noise and to avoid misleading the model during the training.

A single report and its paired image could also contain multiple diseases, enabling multi-label classification. Due to this, labels were made to not be mutually exclusive, allowing the model to detect combinations seen in real-world settings and improving the generalizability of the model across different scenarios.

Moreover, the CheXpert Plus dataset was particularly selected because of its incorporation of patient metadata, including demographics (age, sex, race), BMI measure, and insurance type [10]. This rich metadata supports the development of our model to account for patient-specific factors, considering the imbalance of specific subgroups, and thus allowing for the assessment of model bias and fairness. The dataset incorporates standardized DICOM headers and key clinical observations extracted from text reports with images, allowing for a multimodal design that supports joint reasoning when making diagnoses [10]. The DICOMs and reports comprising the dataset were also de-identified such that they did not contain any protected health information, ensuring confidentiality of the data on which our model was trained [10]. However, a notable limitation of the dataset was its large class imbalance, with the majority of diseases identified as a lung opacity and the minority of diseases identified as pleural other [10]. Furthermore, there existed an imbalance in patient demographics, with the majority of patient data being from White and older individuals [10]. Our model addressed these disparities to minimize the impact of the large imbalance in classes present in the dataset.

### B. Preprocessing Pipeline

We developed a comprehensive preprocessing pipeline that integrated chest radiograph images, patient metadata, and structured labels from radiologist reports to prepare MetaCheX for training. Image data were filtered to include only frontal-view radiographs and were resized and normalized to maintain consistency across samples. Patient metadata, including age, sex, race, and BMI, was extracted from the corresponding DICOM headers and encoded as multihot vectors. These metadata features were processed and concatenated with corresponding image-label pairs, allowing the model to incorporate patient-specific context during training.

To extract structured labels from free-text radiology reports, we used RadGraph, a transformer-based natural language processing (NLP) tool specialized for radiology [11]. Before applying RadGraph, each report underwent tokenization and dependency parsing using spaCy. RadGraph then extracted and categorized observations, either categorizing words that are associated with disease classifications, visual features, or pathophysiologic processes as "Definitely Present," "Uncertain," or "Definitely Absent" [11]. Using relationships between observations and anatomical sites, binary label vectors corresponding to the 14 chest pathologies were created.

While earlier studies, such as Efimovich et al. [12] have employed similar NLP-based pipelines to extract labels from radiology reports, they typically focus only on image and textual modalities. Our proposed pipeline further integrates structured metadata to better reflect how decisions are made in the real world and enhance diagnostic accuracy based on patient-specific conditions. This multimodal preprocessing framework supports the development of a model that is not only more robust but also better equipped to handle unique clinical situations and diverse patient populations.

### C. Model Architecture

MetaCheX integrates patient metadata, model training, and model testing through a hybrid architecture that utilized several different components including a CNN, a multi-layer perceptron, and a shared classification head.

The proposed system implemented an image classification pipeline with a tabular data processing model, allowing MetaCheX to utilize both the chest x-ray and patient demographic data when detecting lung conditions. The architecture involved a pre-trained CNN backbone to extract features from the chest x-ray, a MLP to find patterns in the patient demographic data, and a shared classification head to utilize features extracted from both the chest x-ray and patient demographic data when making the prediction.

Three CNN backbones were employed for training and testing purposes. This allowed for a clearer demonstration of the impact of integrating patient demographic data through an MLP would have on image feature extraction models in general, rather than on just a specific CNN architecture. The three CNN backbones employed were:

- EfficientNet-B3: A convolutional model part of the EfficientNet family, which is known for using a compound scaling method to uniformly scale depth, width, and resolution, allowing for high accuracy with computational efficiency.

- ResNet-50: Another CNN, ResNet-50 is a 50-layer deep residual network that incorporates skip connections to address the vanishing gradient problem, enabling the effective training of deeper architectures.

- VGG-16: The oldest of the 3 architectures, VGG-16 is a deep convolutional architecture composed of 16 weight layers, characterized by its use of very small (3 x 3) convolutional filters and depth.

A multi-layer perceptron was employed in order to extract features from the patient demographic data. An MLP consists of fully connected layers with nonlinear activation functions, and is notable for being able to distinguish data that is not linearly separable. In MetaCheX's MLP, we employed the Swish activation function, which is defined by (1) below, where $x$ is the input to the neuron and $f(x)$ denotes the output of the activation function.

$$f(x) = x * \left(\frac{1}{(1 + e^{-x})}\right) \quad (1)$$

It has been shown to improve performance over traditional functions like ReLU by enabling smoother gradients and better information flow during training. Specifically, the Swish activation function was picked for its ability to handle large, complex datasets with subtle variations and noise, something especially useful in the context of medical datasets.

MetaCheX's MLP consists of two fully connected layers, consisting of dimensions 12x3 and 8x12, before ending with a concatenation layer, which joins together the 8 features from the MLP along with the 1280 features from the CNN (in the case of EfficientNet) before the classifier as shown in Fig. 1.

Fig. 1. Architecture of MetaCheX's metadata branch showing two fully connected layers followed by a concatenation layer that merges the metadata sourced features with image features taken from the CNN.

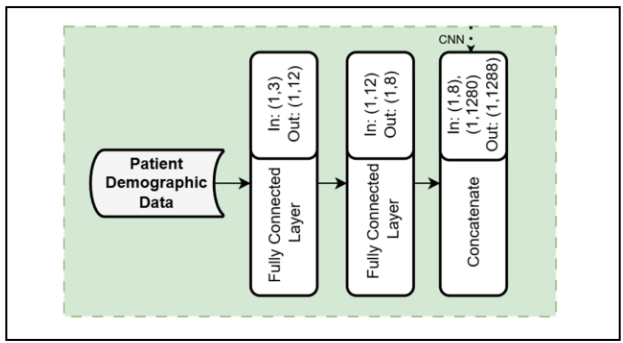

This concatenation layer is essential, allowing for the integration of the features from the patient demographic data to be used alongside the image features when making the classification. The dimensions and number of layers were determined in the training process, as described.

### D. Training Setup

Training was conducted using W&B Sweeps on an RTX 4070 GB. We used Bayesian grid search to determine the number of layers and dimensions of the multi-layer perceptron, learning rate, batch size, and the patient metadata values that would be passed into the multi-layer perceptron. Figure 2 presents a parallel coordinates plot generated during the Weights & Biases hyperparameter sweep for EfficientNet-B3. It illustrates the relationship between the selected hyperparameter — batch size, learning rate, patient demographic features passed into the MLP (meta_features), and the dimensions of the linear layers in the MLP (n_meta_dimensions) — and the resulting AUROC scores obtained during the hyperparameter optimization. Each line corresponds to a unique configuration evaluated in the sweep, and with each color from purple to yellow denoting increasing AUROC values, ranging from 0.774 to 0.796. A grid search was performed on the three backbones, VGG-16, EfficientNet-B3, and ResNet-50, running for 50 epochs each. The best performing model, which used EfficientNet-B3 as a backbone, used patient age, sex, and recent BMI as demographic data to be passed into the model. The MLP had two fully connected linear layers, like previously mentioned, of dimensions 12x3 and 8x12.

Fig 2. Parallel coordinates visualization of EfficientNet-B3 hyperparameter sweep showing AUROC trends across hyperparameters.

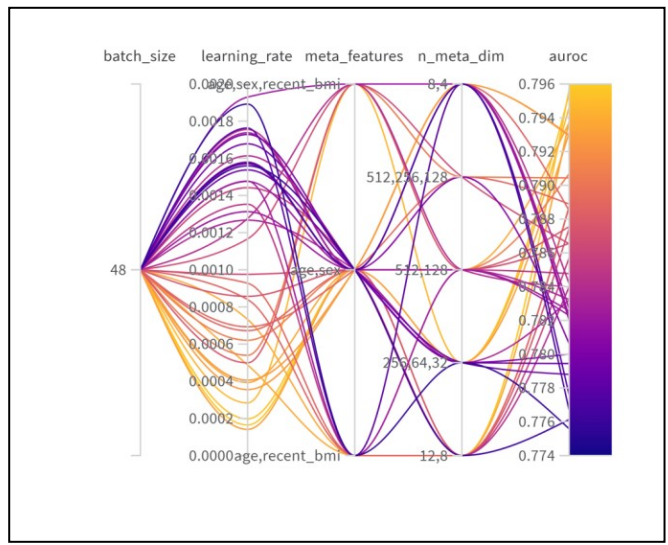

### III. RESULTS

To evaluate the performance of MetaCheX, we conducted experiments using the three CNN backbones: EfficientNet-B3, ResNet-50, and VGG-16. Each backbone was tested with the baseline model, which utilized only chest X-ray radiographs, and our enhanced model, which incorporated patient metadata through an MLP. To assess model performance, the area under the receiver operating characteristic (AUROC) curve was calculated, where values closer to 1 indicate greater effectiveness and diagnostic accuracy. Incorporating metadata into the model consistently improved diagnostic performance across all backbones. Notably, the EfficientNet-B3 model achieved the highest average AUROC, increasing from a baseline score of 0.85538 to 0.88205 with the integration of metadata. Similar trends were observed with ResNet-50 and VGG-16 backbones. ResNet-50 saw an average AUROC improvement from 0.86165 to 0.87998, while VGG-16 improved from 0.85201 to 0.87263 with the integration of metadata, which confirmed that patient-specific contextual information improves general diagnostic performances throughout a range of different neural network architectures.

TABLE I. EFFECT OF METADATA INTEGRATION ON AUROC ACROSS CNN ARCHITECTURES AND PATHOLOGIES

| CNNs | Pathology | | | | | | |
|---|---|---|---|---|---|---|---|
| | *Model* | *Average AUROC* | *Atelectasis* | *Cardiomegaly* | *Consolidation* | *Edema* | *Pleural Effusion* |
| EfficientNet-B3 | Baseline | 0.85538 | 0.79583 | 0.79189 | 0.84671 | 0.90904 | 0.93347 |
| | Added MLP | 0.88205 | 0.82278 | 0.82653 | 0.90171 | 0.91820 | 0.94105 |
| ResNet-50 | Baseline | 0.86165 | 0.81658 | 0.78894 | 0.85832 | 0.91435 | 0.93015 |
| | Added MLP | 0.87998 | 0.85950 | 0.79512 | 0.90386 | 0.92708 | 0.91435 |
| VGG-16 | Baseline | 0.85201 | 0.83158 | 0.81463 | 0.82555 | 0.89183 | 0.89648 |
| | Added MLP | 0.87263 | 0.83016 | 0.84713 | 0.88017 | 0.90738 | 0.89928 |

Beyond overall improvement in average AUROC, the addition of metadata particularly benefited pathologies known to be complex or dependent on patient context. For example, cardiomegaly and consolidation saw significant increases with EfficientNet-B3, from 0.79189 to 0.82653 and 0.84671 to 0.90171, respectively. These improvements suggest that certain diseases may be more reliant on clinical context to achieve higher diagnostic accuracy. These results are displayed in Table 1. However, we also observed that in some instances, such as pleural effusion with ResNet-50, where AUROC declined slightly from 0.93015 to 0.91435, the integration of metadata did not yield improvements. This may reflect that the additional metadata introduced noise or irrelevant signals that would have interfered with the model's diagnoses.

## IV. DISCUSSION

The results demonstrate that incorporating patient metadata alongside imaging can improve diagnostic accuracy across a range of CNN architectures. The consistent improvement in performance across EfficientNet-B3, ResNet-50, and VGG-16 suggests that the benefits of metadata generalize across architectures and serve as meaningful complementary information that clinical models should utilize when making diagnostic decisions. This is particularly important in clinical tasks where different conditions can appear visually similar, or when patient context alters the interpretation of a radiograph image.

Integrating metadata also helps address fairness and bias, which are two issues that show increasing urgency and prevalence in medical AI. Without patient-specific context, models may rely on illusory correlations or overfit to visual patterns that may not be clinically meaningful. By considering metadata, such as age, sex, or BMI, models can make more individualized and equitable diagnoses. This aligns more closely with how human clinicians and radiologists utilize both imaging and patient history to make informed decisions. In this way, our model moves closer to real-world diagnostic reasoning, improving both accuracy and reliability.

While our results show the value of integrating metadata into chest X-ray classification, several limitations remain. First, the impact of metadata varies by disease and architecture. For example, pleural effusion is a condition that shows visible signs in radiograph images, explaining why ResNet-50 performed slightly worse when metadata was added. The integration of metadata into the ResNet-50 architecture may have introduced additional noise or conflicting signals, even when visual evidence of the disease would have already been strong. These findings suggest a need for strategies that selectively weight or filter metadata according to its relevance to the condition being diagnosed in future models.

Second, the metadata used in this study, such as age, sex, race, and BMI, is relatively simple. In real-world settings, however, patient context is richer and more dynamic, including lab results and longitudinal health records. Future work should explore the integration of electronic health record (EHR) data and patient histories to further improve diagnostic accuracy and clinical relevance. Additionally, it's essential to test radiographic machine learning models across diverse environments and institutions that may employ different methods in clinical practice for safe deployment in real-world clinical practice.

Third, the dataset we employed, CheXpert Plus, contains significant imbalance in its distribution of pathology labels and demographic metadata. Imbalances in the dataset, caused by natural prevalence of disease and inherent biases in real-world medical data collection, pose several implications. Models may produce biased predictions and have lower diagnostic accuracy for underrepresented groups. A promising field of research for future work is the use of synthetic data to compensate for these gaps in data. By oversampling underrepresented groups using generative adversarial networks (GANs) or diffusion models to produce realistic chest radiographs, synthetic data is created for the minority class, enhancing generalizability of results. Further benefits of synthetic data include generating visually complex radiographs, which better prepares the model for the real world, where radiograph images may be visually ambiguous.

Our approach proves to be relevant for real-world clinical applications to assist radiologists by providing context-aware support. Especially in high-volume or emergency settings, leveraging a metadata-integrated AI model can help classify cases more accurately and reduce uncertainty in ambiguous radiographs. Our model demonstrates that metadata-embedded systems offer better performance and reduce algorithmic bias to support a safe, trustworthy healthcare environment.

## V. CONCLUSION

We presented MetaCheX, a novel metadata-integrated multimodal model for chest X-ray diagnosis using the CheXpert Plus dataset, demonstrating the significant value of combining structured metadata with imaging features to improve classification performance across multiple CNN backbones. Our best-performing model, EfficientNet-B3 with metadata, achieved an average AUROC score of 0.88205, which is substantially higher than the baseline model without metadata. These results show that metadata provides meaningful context that helps the model make more accurate diagnoses, especially for complex or visually ambiguous images.

MetaCheX also addresses the broader need for fair and context-aware medical AI. By incorporating patient-specific metadata, our model produces more individualized and clinically relevant diagnoses, rather than relying solely on imaging. This reduces the risk of overgeneralization or bias, especially when handling diverse patient populations where radiographs alone may not capture key nuances. MetaCheX also aligns more closely with how clinicians and radiologists reason, considering both imaging data and patient history to inform their decisions. This clinically aligned reasoning framework establishes a new standard for robust and equitable AI-assisted diagnosis. In future work, this metadata-integrated approach may be extended to other medical imaging fields, such as mammography and ultrasonography, where patient context plays a more critical role. Overall, our findings demonstrate the value of metadata in medical machine learning by highlighting its value in enhancing diagnostic performance and improving reliability across diverse patient populations.

## VI. ACKNOWLEDGMENTS

We extend our gratitude to the Stanford Center for Artificial Intelligence in Medicine and Imaging (AIMI) for allowing us to conduct this study using their CheXpert Plus dataset. We also sincerely thank the Stanford University faculty for their guidance and valuable insights in our research.